\documentclass{article}
\usepackage[hyphens]{url}
\usepackage[utf8]{inputenc}
\usepackage{graphicx}
\usepackage{authblk}
\usepackage[english]{babel}
\usepackage{setspace}
\usepackage{fancyhdr}
\usepackage{array}
\usepackage[margin=1in]{geometry}
\usepackage{float}
\usepackage{amsmath}
\usepackage{amssymb}
\usepackage{bbm}
\usepackage{bm}
\usepackage{scalerel,stackengine}
\usepackage[final]{pdfpages}
\usepackage{longtable}
\usepackage{chngcntr}
\usepackage{placeins}
\usepackage{microtype}
\usepackage{tikz}
\usepackage{makecell}
\usepackage{hyperref}
\usepackage{subcaption}
\usepackage{rotating}
\usepackage{soul}
\usepackage{tabularx} 
\usepackage{pdfpages}
\usepackage[square,numbers]{natbib}
\usepackage{RJournal}
\usepackage{amsmath,amssymb,array}
\usepackage{makecell}
\usepackage{booktabs} 
\usepackage{multirow}
\hypersetup{
    colorlinks = true,
    citecolor = {blue},
    urlcolor = {blue},
    menucolor = {blue},
    linkcolor = {blue}
}

\makeatletter
\patchcmd{\@maketitle}{\LARGE \@title}{\fontsize{18}{20}\selectfont\textbf{\@title}}{}{}
\makeatother

\title{Design of Trials with Composite Endpoints with the R Package CompAREdesign}

\author[1]{Jordi Cortés Martinez\thanks{jordi.cortes-martinez@upc.edu}}
\author[2]{Marta Bofill Roig}
\author[1]{Guadalupe Gómez Melis}  

\affil[1]{Universitat Politècnica de Catalunya, Barcelona, Spain} 
\affil[2]{Center for Medical Statistics, Informatics and Intelligent Systems, Medical University of Vienna, Vienna, Austria}

\date{}         
\begin{document}

\maketitle

\begin{abstract}
Composite endpoints are widely used as primary endpoints in clinical trials. Designing trials with time-to-event endpoints can be particularly challenging because the proportional hazard assumption usually does not hold when using a composite endpoint, even when the premise remains true for their components. Consequently, the conventional formulae for sample size calculation do not longer apply. We present the R package \CRANpkg{CompAREdesign} by means of which the key elements of trial designs, such as the sample size and effect sizes, can be computed based on the information on the composite endpoint components. \CRANpkg{CompAREdesign} provides the functions to 
assess the sensitivity and robustness of design calculations to variations in initial values and assumptions. Furthermore, we describe other features of the package, such as functions for the design of trials with binary composite endpoints, and functions to simulate trials with composite endpoints under a wide range of scenarios.
\end{abstract}

%%%%%%%%%%%%%%%%%%%%%%%%%%%%%%%%%%%%%%%%%%%%%%%%%%%%%%%%%%%%%%%%%%%%%%%%%%
%%%%%%%%%%%%%%%%%%%%%%%%%%%%%%%%%%%%%%%%%%%%%%%%%%%%%%%%%%%%%%%%%%%%%%%%%%
%%%%%%%%%%%%%%%%%%%%%%%%%%%%%%%%%%%%%%%%%%%%%%%%%%%%%%%%%%%%%%%%%%%%%%%%%%

\section{Introduction}\label{sect:intro}

Composite endpoints are defined as the occurrence of any of the set of events of interest in trials with binary endpoints and as the time from randomization to the first observed event among all events of interest in time-to-event trials. Composite endpoints are often chosen as primary efficacy endpoints to answer the main research question in confirmatory clinical  trials. 
For instance, progression-free survival, the composite of occurrence of  death and clinical  progression, is one of the most common primary endpoint in oncology clinical trials. Major adverse cardiovascular events (MACE), the composite of cardiovascular death, myocardial infarction, stroke and target vessel revascularization, is frequently used in cardiovascular research \citep{Gomez2014,Freemantle2003,Ferreira-Gonzalez2007,Tomlinson2010}.

There is a large and established literature on designing and analyzing trials with multiple time-to-event endpoints in which case the focus is on the time from randomization to the first observed event among all components. The reader may refer to \cite{Rauch2017} for methods for planning and evaluating clinical trials with primary composite endpoints, \cite{Sozu2015} for sample size determination in trials with with multiple endpoints, and \cite{Ristl2018} for a revision of methods for the analysis of trials with multiple endpoints in small populations. In the particular context of composite endpoints, analyses  that weight the composite's components contribution \citep{Ozga2022,Bakal2015}, matched-paired approaches such as the  win-ratio method \citep{Finkelstein2018,Pocock2012,Dong2019} have been proposed as well as methods for sample size calculation \citep{Cortes2021,Gomez2014,Sugimoto2017}.

To our knowledge, only a few R packages in the R CRAN repository are exclusively focused on composite endpoints. Furthermore, those are more concentrated on the analysis rather than on the design. The \CRANpkg{WR} and \CRANpkg{WWR} packages address the analysis of studies with prioritized composite endpoints through the Win Ratio measure. The \CRANpkg{wcep} package provides the Kaplan-Meier survival curves in the presence of weighted composite endpoints. The \CRANpkg{Wcompo} implements inferential and graphic procedures for the semi-parametric proportional means regression of weighted composite endpoint of recurrent events and death. The \CRANpkg{idem} package implements a procedure for comparing treatments based on the composite endpoint of a functional (unobserved) outcome and a time-to-event endpoint. On the other hand, there are a variety of packages for designing trials with multiple endpoints, but surprisingly do not consider composite endpoints. The \CRANpkg{Mediana} library considers different multivariate distributions to calculate by simulation the sample size needed to analyze multiple events. The \CRANpkg{gMCP} and \CRANpkg{gMCPLite} provide functions for the analysis of trials with multiple hypotheses using graph-based procedures. The \CRANpkg{ADCT} package performs power and sample size calculations for adaptive designs with co-primary endpoints. Finally, the \CRANpkg{cats} allows to simulate platform trials with co-primary binary endpoints.

In this paper, we present the R package \CRANpkg{CompAREdesign} (\cite{CompAREdesign}) which addresses different aspects of the design of randomized controlled trials (RCT) with composite endpoints. This package was conceived as a natural way to provide the functions implemented on the web-app \textit{CompARE} (\url{https://www.grbio.eu/compareCover/}), 
developed with the \CRANpkg{shiny} R package by the same authors as the library presented in this work. 
The main features of the \CRANpkg{CompAREdesign} package are: 1) to anticipate the relative efficiency of the design using a composite endpoint with respect to the design based on a single outcome as primary endpoint; 2) to quantify the expected treatment effect of the intervention on the composite endpoint; and 3) to compute the sample size to detect that treatment effect. Although this paper mainly refers to trials with time-to-event endpoints and presents the R functions for that, the corresponding functions for trials with binary  endpoints are implemented as well in \CRANpkg{CompAREdesign}
We end this article with an overview of the functions of  \CRANpkg{CompAREdesign} for  binary composite endpoints and a brief explanation of some functions to generate data for composite endpoints based on the information of the components.

\section{Methodological background}\label{sect:methods}

Consider a RCT comparing an experimental treatment ($ i=1 $) and a control arm  ($ i=0 $). Suppose that $n^{(i)}$ patients are allocated to arm $i$ and followed for a prespecified time $\tau$. Consider a composite endpoint ($\varepsilon_*$) consisting of two single events, $\varepsilon_1$ and $\varepsilon_2$, and assume that $\varepsilon_1$ is more relevant for the trial purposes. 
Denote by $T_1$ and $T_2$ the times to  $\varepsilon_1$ and  $\varepsilon_2$, respectively, and $T_*$ the time to the composite endpoint $\varepsilon_*$, that is,  $T_*=\min\{T_1,T_2\}$ is the time to the first occurrence of any of the events $\varepsilon_1$ and  $\varepsilon_2$. Furthermore, when required, we denote by $T_k^{(i)}$ the time to $\varepsilon_k$ in the arm $i$.

\subsection{Distribution functions for the composite endpoint}

In order to derive the law for the composite endpoint, we have to distinguish between whether death (or any other fatal event) is included in $\varepsilon_1$ and $\varepsilon_2$. We will refer to case 1: when none of the events includes a fatal event that precludes from observing the other event (e.g., death); cases 2 and 3: when, respectively,  either  $\varepsilon_2$ or $\varepsilon_1$ includes a fatal event; and case 4: when both events include a fatal event. This distinction matters since, depending on the case,  the cause-specific hazard rate function for  $T_1$ and $T_2$ has to be used instead of the corresponding marginal hazard rate functions. For simplicity, in this paper, we will focus on  case 3 ($\varepsilon_1$ including a fatal event). The reader is referred to \cite{Gomez2013} for a thorough explanation of all the cases. 

To derive  the law of $T_*^{(i)}$ the joint distribution  between $T^{(i)}_1$ and $T^{(i)}_2$ ($i=0,1$) has to be characterized. This is accomplished by means of a  copula binding the marginal distributions of $T^{(i)}_1$ and $T^{(i)}_2$ through an association parameter that can been chosen between Spearman's rank correlation coefficient $\rho$ and Kendall's $\tau$.  We assume  that  $\rho$ and $\tau$  are the same in both groups. The marginal laws for $T^{(i)}_k$ ($i=0,1; k=1,2$) are  chosen from the Weibull family of distributions. They depend on a shape parameter which allows increasing, constant  and decreasing hazard functions and a scale parameter that is specified in terms of the probabilities $p_1^{(0)}=p_1^{(0)}(\tau)$ and $p_2^{(0)}=p_2^{(0)}(\tau)$ of	observing endpoints $T_1^{(0)}$ and $T_2^{(0)}$ in the control group.

Finally,  we assume that  treatment groups have proportional (cause-specific) hazard rates for each component and denote by  $\rm{HR}_1$  and $\rm{HR}_2$ the respective (cause-specific) hazard ratios which  have to be anticipated. Without loss of generality, we assume that both events ${\cal E}_1$ and ${\cal E}_2$ are harmful and that the new treatment is expected to reduce the risk of  both events, that is, $HR_k<1$, $k=1,2$.   See \cite{Cortes2021} for further details.

\subsection{Effect size}

In trials with survival endpoints, the efficacy of a treatment is routinely quantified by means of the hazard ratio based on the proportional hazards model. However, this proportionality is not usually met for time-to-event composite endpoints and different summaries, such as the geometric average hazard ratio, the  average hazard ratio, the median ratio or the restricted mean survival time  ratio could be then a more convenient alternative \citep{Rauch2018,Zhao2016}. In what follows, we describe them.

\textbf{The geometric average hazard ratio}, $\rm{gAHR}$, is defined as the exponentiated mean of the logarithm of the hazard ratio, that is,
$
\rm{gAHR}=\exp\left\{{\rm E}(\log  \rm{HR}_*(T)) \right\}
$
where the expectation is taken with respect to a given event-time distribution, which in this case is  the  average distribution of  $T_*^{(0)}$ and  $T_*^{(1)}.$ For a given maximum follow-up time $\tau$, the geometric average hazard ratio  up to $\tau$ is defined as 
\begin{eqnarray} \label{eq:4}
gAHR(\tau)&=& \exp\left\{\frac{\int_0^{\tau} \log\big\{\frac{\lambda_{*}^{(1)}(t)}{\lambda_{*}^{(0)}(t)} \big\} f^{(a)}_*(t) dt}
{p^{(a)}_*(\tau)}  \right\}
\end{eqnarray}
where
$f^{(a)}_*(t)=(f_*^{(0)}(t)+f_*^{(1)}(t))/2$ is  the  average of the density functions  of  $T_*^{(0)}$ and  $T_*^{(1)}$,
$f^{(i)}_*(t)$ is  the density function of $T^{(i)}_* $ ($i=0,1$) and
$p^{(a)}_*(\tau)=(p^{(0)}_*(\tau)+p^{(1)}_*(\tau))/2$
is the average  probability of experiencing the event $\varepsilon_*$ over both groups by time $\tau$.

\noindent In contrast to $HR_*(t)$ which is expected to change over time, $gAHR(\tau)$ is independent of time and keeps its interpretability under non-proportional hazards. Furthermore, the geometric average hazard ratio  and the  all-cause hazard ratios  take identical values under proportionality of the  all-cause hazard rates. Last, $gAHR(\tau)$ is the natural effect measure when using the logrank test to compare the hazard rates of two groups and should be used instead of the standard hazard ratio \citep{Schemper2009}.

\textbf{The  average hazard ratio}, $AHR$, introduced in  \cite{Kalbfleisch1981}, provides a  summary statistic of the effect size that has an interpretation in the absence of proportionality among the hazard rates. The average hazard ratio up to time $\tau$ is defined
\begin{eqnarray} \label{eq:5}
AHR(\tau)&=&\frac{\int_0^{\tau} (\lambda_*^{(1)}(t)/\lambda_*^{(a)}(t))f_*^{(a)}(t)dt}
{\int_0^{\tau} (\lambda_*^{(0)}(t)/\lambda_*^{(a)}(t))f_*^{(a)}(t)dt}
\end{eqnarray}
where
$\lambda^{(a)}_*(t)=\lambda_*^{(0)}(t)+\lambda_*^{(1)}(t)$ is the overall hazard function among the two groups. The average hazard ratio up to time $\tau$ can be interpreted as an average of the hazard ratios at all death times, also under non-proportional hazards.

\noindent Previous work based on simulations (\cite{Cortes2021}) shows  that the values of  $AHR$ and  $gAHR$ are very close, which entails that $gAHR$  could be interpreted as a measure of proportional hazards just in the same way as $AHR$. In addition, the $gAHR$ has the advantage of a direct relationship with the sample size calculation (as explained below).

\textbf{The median ratio}, $mR_*$,  corresponds to the ratio of the median times to  $\varepsilon_*$ over both arms. Therefore, if
$m_*^{(i)}=inf\{t: S_*^{(i)}(t)<0.5\}$ where  $S_*^{(i)}(t)$ is  the survival function of the composite endpoint in group $i$ ($i=0,1$), then
\begin{eqnarray} \label{eq:6}
mR_*&=&\frac{m_*^{(1)}}{m_*^{(0)}}.
\end{eqnarray}

\noindent The median ratio is another appropriate alternative to the $gAHR(\tau)$ and the $AHR(\tau)$, which coincides with those  in case the event rate is constant over time. Furthermore, the  $mR_*$ gives a measure of the time-to-event gain in one group relative to another, giving it greater interpretability than risk-based measures (e.g., $gAHR(\tau)$ or $AHR(\tau)$) (\cite{Cortes2014}).

\textbf{The restricted mean survival time  ratio},  $RMSTR_*(\tau)$,  corresponds to the ratio of the restricted mean survival times to  $\varepsilon_*$ over both groups up to time $\tau$. The restricted mean survival time (RMST)  of $\varepsilon_*$ in arm $i$ up to time $\tau$ (\cite{Royston2013}) is the area under the survival curve up to $\tau$, given by $RMST_*^{(i)}(\tau)=\int_0^{\tau}S_*^{(i)}(t)dt$.
The restricted mean survival time  ratio is then defined as follows:
\begin{eqnarray} \label{eq:7}
RMSTR_*(\tau)&=&\frac{RMST_*^{(1)}(\tau)}{RMST_*^{(0)}(\tau)}.
\end{eqnarray}

\noindent Although the difference in restricted means as an alternative measure of treatment effect is often used, we advocate here for the ratio for analogy with all the other effect size measures.

\subsection{Sample size}

Sample size calculation to detect a hypothesized difference between treatments is a key point in the design of a RCT. In survival trials with composite endpoints, the  sample size can be based on the geometric average hazard ratio, $gAHR$, in case the proportional hazards assumption can be assumed to hold for the components, but not for the composite endpoint. The required number of events, sample size and power formulae are based on the non-centrality parameter of the logrank test under the alternative hypothesis which is a function of the $gAHR$. 

Suppose we aim at testing the superiority of the new treatment $(i=1)$ against the  control arm and that the logrank test  statistic $Z_*$   is used for the null hypothesis of no effect on $T_*$. If using the geometric average hazard ratio  as the treatment effect measure, the null hypothesis of no effect on $T_*$ will be rejected for a one-sided $\alpha$ significance level whenever $Z_*<-z_{\alpha}$,  where $z_{\alpha}$ is the $\alpha$-quantile of the standard normal distribution. Note here that negative values of $Z_*$ favor the new treatment. Since  $Z_*$ follows a normal distribution with mean $\mu_*(\tau)$ and variance 1, the power $1-\beta$ is such that $1-\beta={\rm Prob}\{Z_*<-z_{\alpha}\}$. Hence,
the total sample size for both groups 
for a balanced design (equal sample size in both groups) is:
\begin{eqnarray}\label{nSS}
n&=&\frac{4(z_{\alpha}+z_{\beta})^2}{p_*^{(a)}(\tau)  \left(\log(gAHR(\tau)\right)^2}
\end{eqnarray}
and the expected number of composite endpoint events $e_*=n\cdot p_*^{(a)}(\tau)$ is
given by
\begin{eqnarray}\label{events}
e_*&=&\frac{4(z_{\alpha}+z_{\beta})^2}{\left(\log(gAHR(\tau)\right)^2} 
\end{eqnarray}

To obtain expression \eqref{events} from expression \eqref{nSS} (or vice versa), the same follow-up period ($\tau$) had to be assumed for all study participants, regardless of how recruitment was carried out. Observe that, formula \eqref{events} corresponds to Schoenfeld's formula \citep{SCHOENFELD1981} for the required number of events if the hazard ratio is substituted by the geometric average hazard ratio.  The main difference lies, however, in that while in Schoenfeld's formula you are assuming that the hazard rates are proportional when dealing with a composite endpoint the all-cause hazard rates do not have to be proportional. 

\subsection{Endpoint selection}

When designing a trial with multiple endpoints, one might wonder which endpoint is better from a point of view of the statistical efficiency of the trial. Specially when composite endpoints are considered, adding more endpoints to the primary composite endpoint can dilute the effect of some of the most relevant endpoints. The Asymptotic Relative Efficiency (ARE) was proposed as a measure to evaluate the statistical efficiency gain of the composite endpoint versus one of its components. Specifically, the ARE for survival composite endpoints compares the efficiency of using the logrank test $Z_*$ based on the composite endpoint  $\varepsilon_*$ versus the logrank test $Z$ based on the relevant endpoint  $\varepsilon_1$. In what follows, we sketch the main idea of the method, but for further details, we refer to the article \citep{Gomez2013}.

Given that both tests $Z$ and $Z_*$ are asymptotically N(0,1) under $H_0$: no effect on $T_1$ and $H_0^*$: no effect on $T_*$ and are asymptotically normal with variance 1 under a sequence of contiguous alternatives to the null hypothesis, the ARE is given by the square of the ratio of their non-centrality parameters $\mu$ and $\mu_*$, respectively, and admits  the following expression
\begin{equation}\label{AREcase1}
{\rm ARE}(Z_*, Z)=\left(\frac{ \mu_*}{\mu}\right)^2=
\frac{\left(
	\int_0^{1}
	\log\big\{\lambda_{*}^{(1)}(t)/\lambda_{*}^{(0)}(t) \big\}
	f_*^{(0)}(t)  dt\right)^2
}{
	(\log{\rm HR_1})^2
	(\int_0^{1}  f_*^{(0)}(t)  dt)
	( \int_0^{1} f^{(0)}_1(t)  dt)
}
\end{equation}
where  $f_1^{(0)}(t)$ and $f_*^{(0)}(t)$ correspond to the densities, under the control group, of $T_1$ and $T_*$, respectively, $\lambda_{*}^{(0)}(t)$ and $\lambda_{*}^{(1)}(t)$ correspond to the hazard functions of $T_*$ under the control and experimental groups, respectively and ${\rm HR_1}$ stands for the constant hazard ratio for endpoint $T_1$. The ARE measure can be roughly interpreted as the ratio of the required sample sizes using $\varepsilon_1$ versus $\varepsilon_*$ to attain the same power for a given significance level. This measure therefore yields to the following criterion: whenever ARE $>$1, choose $\varepsilon_*$ as primary endpoint to guide the study; otherwise, use $\varepsilon_1$. 

\section{Overview of CompAREdesign R package}\label{sect:comparedesign} 

In this section, we provide a general description of the package. We start explaining the installation and dependencies, and continue describing the functions and arguments for time-to-event functions. We postpone to section \ref{sect:binary} the description of the functions for composite binary endpoints. \CRANpkg{CompAREdesign} provides either numerical and graphical outputs for all methods described in the methodological section for the design of trials with composite endpoints. Further details on the usage of the functions can be found in the corresponding R package manual.

\subsection{Installation and dependencies}

The \CRANpkg{CompAREdesign} package is available on CRAN at:
\begin{center}
	\url{https://cran.r-project.org/web/packages/CompAREdesign/index.html}.
\end{center} 
and can be installed and loaded by running the R commands:
\begin{example*}
	> install.packages("CompAREdesign")
	> library("CompAREdesign")
\end{example*}

The package depends on the \CRANpkg{copula} package \citep{copula2022}, which implements joint distributions binned by copulas; the packages \CRANpkg{ggplot2} \citep{ggplot22016} and \CRANpkg{ggpubr} \citep{ggpubr2020},  needed for the graphical tools; and  the packages \CRANpkg{rootSolve} \citep{rootSolve2009} and   \CRANpkg{numDeriv} \citep{numDeriv2019}, required to numerically compute some integrals needed in the developments. 

\subsection{Explanation of functions} 

\CRANpkg{CompAREdesign} consists of thirteen functions. Six of them refer to time to event composite endpoints, and seven refer  to the binary composite endpoint. The functions whose name ends with \texttt{tte} are those for the time-to-event case, \texttt{cbe} for the binary case, and two extra functions named \texttt{lower\_corr} and \texttt{upper\_corr} concern the bounds of the correlation between binary endpoints. All these functions are implemented in the shiny web-tool \textbf{CompARE} allowing  an interactive way of using them. Table \ref{table:compare_functions} gives a high-level description of these functions and relates them to the capabilities of the app. 

\begin{table}[!ht]
	\centering \scriptsize
	\caption{R functions included in \CRANpkg{CompAREdesign} package along with the corresponding description and the CompARE web-tool's tab where the function is used.}
	\label{table:compare_functions}
	\resizebox{\textwidth}{!}{
		\begin{tabular}{ccc}
			\hline 
			\multicolumn{3}{l}{ \textbf{Functions for the composite of time-to-event endpoints}} \\ [2mm]
			\hline \\

			\textbf{R function} &  \textbf{Description} &  \textbf{CompARE web-tool tab} \\ [2mm]
			
			\hline
			\texttt{surv\_tte} & Computes the survival function for the& Summary\\   
			& composite endpoint and both components &  \\ [2mm] 
			
			\texttt{effectsize\_tte} & Computes the treatment effect  & Effect size \\   
			& for the composite endpoint &  \\ [2mm]  
			\texttt{samplesize\_tte}  & Computes the sample size & Sample size \\  
			& for the composite endpoint  &  \\ [2mm]   
			\texttt{ARE\_tte}  & Computes the ARE method for   & Endpoint selection \\ 
			& time-to-event endpoints & \\ [2mm]
			\texttt{plot\_tte}  & Returns four plots related  &  All tabs\\ 
			& to previous features & \\ [2mm]
			\texttt{simula\_tte}  & Simulates time-to-event data for the  &  (Not implemented) \\ 
			& composite and its components. & \\ [2mm]
			
			\hline 
			\multicolumn{3}{l}{\textbf{Functions for the composite of binary endpoints}} \\ [2mm]
			\hline \\
			\textbf{R function} &  \textbf{Description} &  \textbf{CompARE web-tool tab} \\ [2mm]
			
			\hline
			\texttt{prob\_cbe} &  Computes the probability  & Summary \\    
			&   of observing the composite &   \\ [2mm]  
			\texttt{lower\_corr} & Computes the lower limit  & \multirow{2}{*}{Association Measures} \\  
			& for Pearson's correlation &  \\ [2mm]  
			\texttt{upper\_corr} & Computes the upper limit & \multirow{2}{*}{Association Measures} \\  
			& for Pearson's correlation &  \\ [2mm] 
			\texttt{effectsize\_cbe} & Computes the expected treatment & Effect size \\    
			&  effect for the composite endpoint  &  \\ [2mm]  
			\texttt{samplesize\_cbe}  & Computes the needed sample size & Sample size \\  
			& for the composite endpoint  &  \\ [2mm]   
			\texttt{ARE\_cbe}  & Computes the ARE method for    & Endpoint binary endpoints \\ 
			& composite  endpoint & \\ [2mm]
			\texttt{simula\_cbe}  & Simulates binary data for the  &  (Not implemented) \\ 
			& composite and its components. & \\ [2mm]
			
			\hline
			
	\end{tabular}}
\end{table} 

In what follows we describe the time-to-event functions included in the package:

\begin{itemize}
	\item \textbf{\texttt{surv\_tte}} draws, in the same graphical window, for each treatment arm, the survival functions of each component as well as of the composite endpoint.
	\item \textbf{\texttt{effectsize\_tte}} provides the anticipated treatment effect for the composite endpoint in terms of the following measures: the geometric average hazard ratio $gAHR_*(\tau$), the average hazard ratio $AHR_*(\tau$), the median ratio $mR_*$, and the restricted mean survival time ratio $RMSTR_*(\tau$). In addition, for each treatment arm, this function returns: the $RMST_*^{(i)}$, the $m_*^{(i)}$ and the probability of observing the event in each group, $p_*^{(i)}$ ($i=0,1$).
	\item \textbf{\texttt{samplesize\_tte}} returns the required sample size for the three following designs which depend on which primary endpoint is used: 1) sample size for a design based on the relevant component of the composite endpoint $\varepsilon_1$; 2) sample size for a design based on the second component $\varepsilon_2$; and 3) sample size for a design based on the composite endpoint $\varepsilon_*$.
	\item \textbf{\texttt{ARE\_tte}} provides the Asymptotic Relatively Efficiency (ARE) value for comparing the efficiency of using a design with the composite endpoint as primary endpoint versus a  design with the first component as the primary endpoint. An ARE value larger than one indicates a benefit, in terms of power efficiency, when using a composite endpoint and otherwise, the relevant endpoint is the preferred one. 
	\item \textbf{\texttt{plot\_tte}} is a summary function that draws the following four plots: 1) the survival function of the composite endpoint for each treatment arm; 2) the expected hazard ratio over the follow-up time; 3) the ARE value with respect to  the correlation between components; and 4) the required sample size as function of the correlation. 
	\item \textbf{\texttt{simula\_tte}} simulates two-arm trials with composite endpoints. This functions is further explained in Section \ref{sect:sim}.
\end{itemize} 

In the next section, we describe the arguments needed for these functions, and we illustrate and further explain the functions in Section \ref{sect:comparette}. 

\subsection{Explanation of the parameters}

Most functions in \CRANpkg{CompAREdesign} package use common arguments. All arguments in \CRANpkg{CompAREdesign} are briefly described in Table \ref{table:compare_arguments}. Note that in some cases, the same arguments are used for time-to-event and binary functions. 

\begin{table}[!ht]
	\centering
	\caption{Arguments of the functions for binary (\textit{B}) and time-to-event (\textit{T}) endpoints and their description}
	\label{table:compare_arguments}
	\resizebox{\textwidth}{!}{\begin{tabular}{clcc}
			\hline
			\textbf{Argument} &  \textbf{Description} & B & T \\ [2mm]
			\hline 
			\texttt{p0\_e1}, \texttt{p0\_e2} &	
			Probability of occurrence of $\varepsilon_1$ and  $\varepsilon_2$ in the control arm  (Numeric) & X & X
			\\
			\texttt{eff\_e1}, \texttt{eff\_e2} &	
			Anticipated effect for the composite component $\varepsilon_1$ and  $\varepsilon_2$ (Numeric) & X & 
			\\
			\makecell{\texttt{effm\_e1}, \texttt{effm\_e2}, \\ \texttt{effm\_ce}}&	
			Effect measure used for the event $\varepsilon_1$, $\varepsilon_2$ and $\varepsilon_*$  (Character)  & X &
			\\
			\texttt{HR\_e1}, \texttt{HR\_e2}  &	
			Expected cause specific hazard ratio for the $\varepsilon_1$ and $\varepsilon_2$ (Numeric) &  & X 
			\\
			\texttt{beta\_e1}, \texttt{beta\_e2} &	
			Shape parameter of a Weibull distribution for the $\varepsilon_1$, $\varepsilon_2$ in the control arm  (Numeric) &  & X
			\\
			\texttt{rho} &	
			Pearson, Spearman or Kendall correlation between $\varepsilon_1$ and $\varepsilon_2$  (Numeric)  & X & X
			\\
			\texttt{rho\_type} & Type of correlation (Character)  & X & 
			\\
			\texttt{case} & 1 to 4 depending on whether death is included in $\varepsilon_1$ or $\varepsilon_2$ or both or neither (Numeric)  & X & \\
			\texttt{copula} & Type of copula to build the joint distribution (Character)  & X & 
			\\
			\texttt{alpha}  & Probability of Type I error (Numeric) & X & X 
			\\
			\texttt{beta} & Probability of Type II error (Numeric) & X & 
			\\
			\texttt{power} & Power to detect a real treatment effect on composite endpoint (Numeric) & & X 
			\\
			\texttt{ss\_formula} & Formula for the sample size calculation on the single components (Character) & & X 
			\\
			\texttt{unpooled} & Variance estimate used for the treatment effect (Character)  & X &
			\\ 
			\texttt{followup\_time} & Time of follow-up ($\tau$) (Numeric)  &  & X
			\\
			\texttt{subdivisions} & Number of points to perform numerical calculations  (Integer)  &  & X
			\\
			\texttt{plot\_res} & Indicates if the plot should be displayed (Logical)  &  & X
			\\
			\texttt{plot\_store} & Indicates if the plot should be stored (Logical)  &  & X
			\\
			\texttt{sample\_size} & Desired sample size for each arm when simulating data (Integer)  &  & X
			\\ 
			\hline 
	\end{tabular}}
\end{table}

\textbf{\texttt{p0\_e1}} and \textbf{\texttt{p0\_e2}} represent the probabilities of observing the event in the reference arm during the follow-up period. It must be kept in mind that, for a single participant, \texttt{p0\_e1} is the probability of observing the $\varepsilon_1$ even if $\varepsilon_2$ had previously observed and vice versa. Those probabilities could be easily obtained from the literature, for example,  using the proportion of observed events at the end of the study in trials that the pertinent component is used as primary endpoint.  

\textbf{\texttt{HR\_e1}} and \textbf{\texttt{HR\_e2}} are the anticipated treatment effects for the composite components $\varepsilon_1$ and $\varepsilon_2$ in terms of the cause-specific hazard ratios. They could also be found in previous published trial results with $\varepsilon_1$ or $\varepsilon_2$ as primary endpoints.

The arguments \textbf{beta\_1} ($\beta_1$) and \textbf{beta\_2} ($\beta_2$) are the shape parameter of the marginal Weibull distribution behind each component. They can be guessed taking into account that values below unity imply a decreasing risk of that event occurring over time and vice versa, while a $\beta_j=1$, $j=1,2$ indicates a constant risk. Our recommendation is:

\begin{itemize}
	\item If the risk of suffering the event decreases throughout the follow-up period (e.g., an infection after a surgical intervention), then assign a $\beta_j=0.5$
	\item If the risk of suffering the event remains constant during the follow-up period (e.g., death after a non-invasive intervention in mild patients), then assign a $\beta_j=1$
	\item If the risk of suffering the event increases during the follow-up period (e.g., contracting an infectious disease at the start of a pandemic), then assign a $\beta_j=2$
\end{itemize}

The ARE method is based on a copula binding the distribution of the  times to $\varepsilon_1$ and $\varepsilon_2$ through a measure of association, for example, the Spearman’s rank correlation coefficient (\textbf{\texttt{rho}}) between these times.  As it can be hard to anticipate a measure of association, the package provides both the sample size and the ARE for different positive correlation values (we assume that the association between components cannot be negative). Time-to-event functions implement another association measure, the Kendall's $\tau$, that could be used instead of the Spearman’s $\rho$ by means of the argument \textbf{\texttt{rho\_type}}. A note of caution is that  Spearman’s $\rho$ cannot be obtained from the data in the presence of competing risk, i.e. when $\varepsilon_1$ and/or $\varepsilon_2$ is death.

The \textbf{\texttt{case}} parameter indicates in which of the two components that make up the composite event, a fatal event is present. This is important to address the scenario of competing risks:
\begin{itemize}
	\item If \code{case = 1}, then none of the events of interest includes a fatal event that precludes from observing other events (such as death)
	\item If \code{case = 2}, then $\varepsilon_2$ is a fatal event
	\item If \code{case = 3}, then $\varepsilon_1$ is a fatal event
	\item If \code{case = 4}, then both components are fatal events. An example of this is cause-specific mortality, such as the composite includes deaths by different causes (e.g., death from heart disease, death from cancer, death from other causes, etc.)
\end{itemize}

The \textbf{\texttt{copula}} argument indicates which type of copula is used to obtain the joint distribution. Currently, the Archimedean copulas of Frank (default), Gumbel and Clayton are implemented. The former provides the same weight to all the events along time, while the Gumbel and Clayton copulas provide more weight at the start and at the end of the follow up, respectively \citep{Trivedi2007}.

\textbf{\texttt{alpha}} and \textbf{\texttt{power}} are relevant parameters for the sample size calculation. \textbf{\texttt{alpha}} ($\alpha$) represents the probability of Type I error, that is the probability of finding a statistically significant treatment effect when it really does not exist. The \textbf{\texttt{power}} ($1-\beta$) is the desired probability of detecting a treatment effect when it really exists. The parameter \textbf{\texttt{ss\_formula}} gives the user two options to calculate the required number of events: the Schoenfeld's or the Freedman's formula (see \cite{Friedman1996}).

\textbf{\texttt{followup\_time}} ($\tau$) argument represents the length of the follow up (measured in any time unit). It facilitates the interpretation of some graphics since the range of the x-axis is fitted to this follow-up period. Moreover, time-dependent effect measures, such as the $mR_*$ or the $RMST_*$ are calculated taking into account this value. By default, an unitary time is assumed (e.g., one year).

The purpose of the parameter \textbf{\texttt{subdivisions}} is to set in how many points some functions (e.g., $\text{HR}^*$) are evaluated to be plotted. The higher the value, the more accuracy at a higher computational cost.

Finally, \textbf{\texttt{plot\_res}} and \textbf{\texttt{plot\_store}} indicate if the plots returned by some functions should be displayed and stored (for further customization), respectively.

\section{CompAREdesign for time-to-event endpoints}\label{sect:comparette}

This section introduces the main features of the package via a lung cancer case study. This illustration exemplifies those aspects of the design of a RCT that can benefit of the capabilities of   the  implemented functions in this package. 

\subsection{Planning a lung cancer trial based on ZODIAC trial} \label{case_study}

We show a case study to design an RCT based on the results from the ZODIAC trial (\cite{Herbst2010}), which compared the efficacy of vandetanib plus docetaxel versus placebo plus docetaxel as second-line treatment in patients with advanced non-small-cell lung cancer. The co-primary endpoints of this trial were overall survival (OS, $\varepsilon_1$) and progression free survival (PFS, $\varepsilon_{*}$), defined as the absence of death and disease progression (DP, $\varepsilon_2$). Let's assume that we want to conduct a new RCT to compare two treatments similar to those tested in the ZODIAC trial. 

Some of the required values for the arguments of the functions are obtained from the published article of the trial. It reports: i) the cause-specific HRs for each component: 0.91 for OS ($\varepsilon_1$) and 0.77 for DP ($\varepsilon_2$); ii)  the probabilities of observing deaths (taken into account those subsequent to the DP) or DP were 0.59 and 0.74, respectively. However, some assumptions have to be made because there is no  other further information on the components. For example, we can expect that the risk of death remains constant ($\beta_1=1$) over time while DP has an increasing risk ($\beta_1=2$) to be occurred. Also, we could anticipate a moderate correlation ($\rho=0.5$) between both components. The first step before using the package is to specify all the arguments on which we are going to base our RCT design: 

\begin{example}
	> p0_e1         = 0.59
	> p0_e2         = 0.74
	> HR_e1         = 0.91
	> HR_e2         = 0.77
	> beta_e1       = 1
	> beta_e2       = 2
	> case          = 3
	> copula        = 'Frank'
	> rho           = 0.5
	> rho_type      = 'Spearman'
	> followup_time = 1
	> alpha         = 0.05
	> power         = 0.80
	> ss_formula    = 'schoenfeld'
\end{example}

\subsection{General overview and characterization of the law of \texorpdfstring{$T_*$}{Lg}}

\begin{figure}[!ht]
	\centering
	\resizebox*{13.5cm}{!}{\includegraphics[width=1\linewidth]{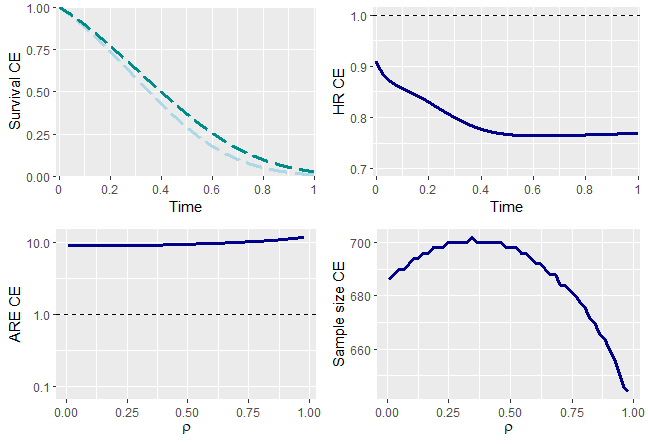}}
	\caption{\textbf{Top left}: survival curves of the composite endpoint for both treatment arms. \textbf{Top right}: anticipated $\rm{HR}_*(t)$ over time. \textbf{Bottom left}: asymptotic relative efficiency as a function of the correlation between components. \textbf{Bottom-right}: required sample size depending on the correlation between components.}
	\label{Figure_ZODIAC}
\end{figure}

The \code{plot\_tte} function is intended to be an all-in-one function that displays all relevant plots for decision making in an RCT. Figure \ref{Figure_ZODIAC} shows the main graphics to assess the behavior of the CE in the trial design.

\begin{example}
	> plot_tte(p0_e1, p0_e2, HR_e1, HR_e2, beta_e1, beta_e2, case, copula, rho, rho_type, 
	followup_time, alpha, power, ss_formula)
\end{example}

The two top plots in  Figure \ref{Figure_ZODIAC} show the survival and  hazard ratio functions of the composite endpoint over time. The survival curves (top-left) reveal that almost no censored data is expected in either of the two treatment arms due to the high proportion of expected events at the end of the study (0.59 and 0.74 for $\varepsilon_1$ and $\varepsilon_2$, respectively). If one wants to obtain, in addition, the survival curves for each component, the function \code{surv\_tte} can be called for this purpose using the same input parameters that \code{plot\_tte}. The follow-up period can be specified to plot the survival curves at the relevant scale for a better interpretation. For instance, we can indicate that we plan to follow the participants during two years by the \code{followup\_time} argument:

\begin{example}
	> surv_tte(p0_e1, p0_e2, HR_e1, HR_e2, beta_e1, beta_e2, case, copula, rho, 
	rho_type, plot_res = TRUE, plot_store = FALSE, followup_time = 2)
\end{example}

On the top-right of Figure \ref{Figure_ZODIAC}, the hazard ratio of the composite endpoint  shows that does not remain constant over time. Indeed, $HR_*(t)$ decreases from $HR_*(0)\approx 0.90$ to $HR_*(0.5)\approx 0.77$ in the first half of the study. From here on, the treatment effect, given by the hazard ratio, seems to remain constant for the rest of the follow up. This plot can be very useful when planning a trial to visually assess whether the proportional hazards assumption is satisfied in a specific setting.

As the association between the composite components, in the planning phase of a trial, is commonly unknown, our package provides the ARE and the sample size as a function of the selected association measure  (see Figure \ref{Figure_ZODIAC} with Spearman's rank correlation $\rho$). The bottom-left  plot allows  the user to  assess the possible impact that the  correlation has on the ARE  value. 

Note that, for this particular example, the design that includes the composite endpoint as the primary endpoint is clearly more efficient than the one that considers death ($\varepsilon_1$) as the main outcome of interest. This holds regardless of the correlation between components, as the ARE takes values around 10 for all possible values  of Spearman's correlation. 
Therefore, in this situation, the value of $\rho$ should not guide our decision. 

On the other hand, on the bottom-right plot, we can observe the sample size of the composite endpoint in terms of the correlation  $\rho$. In this particular case, the sample size takes values between 650 and 700, not varying greatly on basis of the correlation. A conservative choice could be to use the largest sample size (700 in this case) in this trial.

\subsection{Effect size calculation by means of \texttt{effectsize\_tte}}

As mentioned in the methodological section, the common effect measure in survival trials is the hazard ratio which relies on the proportional hazards assumption. Other effect measures have been proposed instead, which might be useful when the hazard ratio deviates from the proportional hazards assumption. The function \texttt{effectsize\_tte} allows to anticipate the expected treatment effect for the composite endpoint based on the information of the components. The function returns the treatment effect is given using different measures, as well as a plot of the $HR_*(t)$ over time. 

To compute the anticipated effect for the hypothetical trial based on the ZODIAC trial, we can use then:
\begin{example}
	> effectsize_tte(p0_e1, p0_e2, HR_e1, HR_e2, beta_e1, beta_e2, case, copula, rho, 
	rho_type, followup_time = 4, subdivisions = 1000, plot_res = FALSE, 
	plot_store = FALSE)
\end{example}

\begin{example*}
	Effect measure Effect value | Group measure Reference Treated
	-------------- ------------ | ------------- --------- -------
	gAHR           0.7989       |                                
	AHR            0.7990       |                                
	RMST ratio     1.1270       | RMST          1.5143    1.7066 
	Median ratio   1.1323       | Median        1.4167    1.6042 
	| Prob. E1      0.5900    0.5557 
	| Prob. E2      0.7400    0.7433 
	| Prob. CE      0.9896    0.9712 
\end{example*}

The output gives two summary measures of the $HR_*(t)$ over time: the geometric average hazard ratio  $gAHR_*(\tau)$ (see \eqref{eq:4}) and the average hazard ratio  $AHR_*(\tau)$ (see \eqref{eq:5}), which in this case provide quite similar values (both, close to 0.80, as one can deduce from the second plot of Figure \ref{Figure_ZODIAC}). When $HR(t)$ is not constant, the $gAHR_*(\tau)$ and the $AHR_*(\tau)$ should be cautiously interpreted since they do not reflect an overall effect. The function also reports other effect measures:  the restricted mean survival time up to the end of follow-up ($\tau$)  $RMSTR_*(\tau)$ (see \eqref{eq:7}), and the median ratio $mR_*(\tau)$(see \eqref{eq:6}). In this case, these measures  take values about $1.13$, indicating that the intervention provides a 13\% gain in the time to event of the $\varepsilon_*$. Finally, some measures for each treatment arm are provided, such as the $RMST$, the median, and the probability of observing the events ($\varepsilon_1$, $\varepsilon_2$, or $\varepsilon_*$). The output also includes the plot of the $HR_*(t)$ (not shown here).

The medians for the composite endpoint in each arm can always be calculated using the marginal Weibull distributions. However, the results should be taken cautiously when some of these medians go beyond the follow-up time ($\tau$).

\subsection{Sample size calculation by means of \texttt{samplesize\_tte}}

We have implemented the sample size calculation for survival trials with composite endpoints in the function \code{samplesize\_tte}. This function allows the user to compute the required sample size to have $1-\beta$ power to detect the treatment effect at  significance level $\alpha$. The function \code{samplesize\_tte} returns the total sample size (assuming equal-sized arms) that would be required if the trial is designed to detect an effect on the composite endpoint. For comparison, the needed sample sizes for trials using 
endpoint 1 ($\varepsilon_1$) or endpoint 2 ($\varepsilon_1$) as primary endpoint instead, are also provided.

We show the usage of the function using the ZODIAC trial as an example. We use the values displayed in Section \ref{case_study} as input arguments, and additionally we set the power and significance level to be equal to 0.80 and 0.05, respectively.
The usage of \code{samplesize\_tte} is then as follows:
\begin{example}
	> samplesize_tte(p0_e1, p0_e2, HR_e1, HR_e2, beta_e1, beta_e2, case, copula, rho, 
	rho_type, alpha, power, ss_formula)
\end{example}

\begin{example*}
	Endpoint           Total sample size
	--------           -----------------
	Endpoint 1         6162             
	Endpoint 2         620              
	Composite endpoint 636
\end{example*}
Note that the design that requires a smaller sample size is the one that considers DP ($\varepsilon_2$) as the primary endpoint. This is due to the fact that the effect for  endpoint 2 is larger than the one for endpoint 1 and therefore it requires less participants to have 0.80 probability to detect the effect. Furthermore, the number of expected observed events is higher for the DP (0.74 versus 0.59), which also makes the sample size for the design that only considers endpoint 1 ($\varepsilon_1$, OS) to be larger than the one using endpoint 2. The consideration of the composite endpoint as primary endpoint leads to a reduction of  the sample size up to almost a 90\% in comparison of using the design that considers $\varepsilon_1$ as the only relevant event.

\begin{figure}[!ht]
	\centering
	\resizebox*{13.5cm}{!}{\includegraphics[width=1\linewidth]{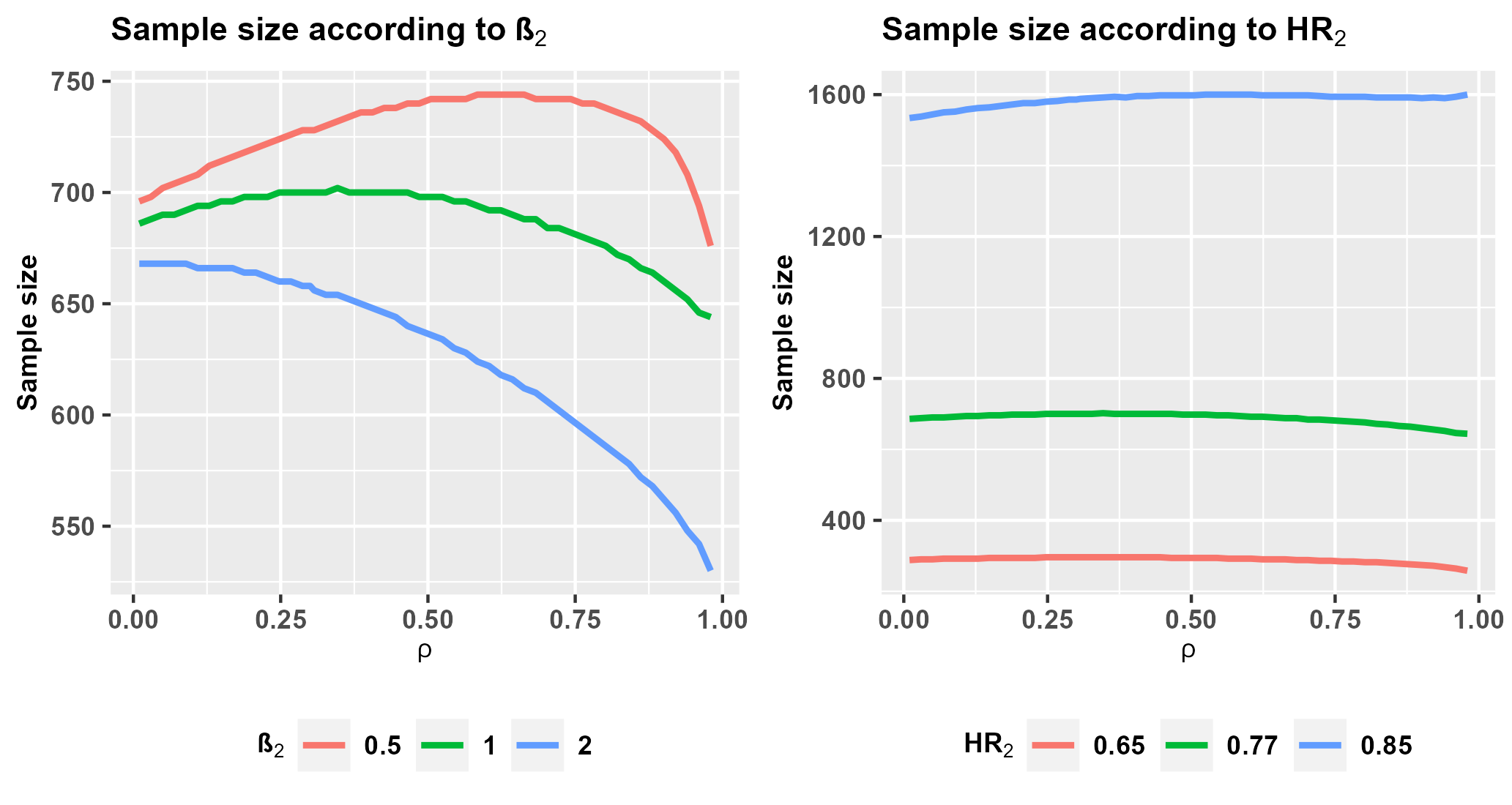}}
	\caption{Sample size under several scenarios. At left, sample size stratified according to the shape parameter $\beta_2$ of $\varepsilon_2$: decreasing risk over time (red, $\beta_2=0.5$); constant risk (green, $\beta_2=1$); and increasing risk (blue, $\beta_2=2$). At right, sample size stratified according to the  hazard ratio $\text{HR}_2$ of $\varepsilon_2$: large treatment effect (red, $\text{HR}_2=0.65$); moderate treatment effect (green, $\text{HR}_2=0.77$); low treatment effect (blue, $\text{HR}_2=0.85$). In both plots, values are represented over different values of Spearman's rank correlation coefficient $\rho$.}
	\label{Figure_samplesize}
\end{figure}

As it is well known the sample size calculation is in general sensitive to modification of the parameters this relies on. 
The choice of certain input arguments in the \code{samplesize\_tte} function might significantly modify the obtained sample size. CompAREdesign can be useful to evaluate the robustness or sensitivity of the sample size calculations according to the assumptions and parameter inputs. As an example, 
Figure \ref{Figure_samplesize} shows the sample size calculation  according to the Weibull shape parameter ($\beta_2$) and the treatment effect ($HR_2$) for the second component. If there is low or no association between both components, the choice of $\beta_2$ barely impacts on the total sample size, which ranges between 670 and 700. The discrepancies are more extensive as the correlation increases. For example, for Spearman's correlation equal to 0.75, the total size could range from 600 if the DP risk increases over time to 750 if the hazard decreases. Obviously, modifying the treatment effect is more critical; the right graph shows that considering an HR of 0.85 instead of 0.65 can increase the total sample size from around 300 to 1,600. These plots have been obtained by setting the parameter \code{plot\_store = TRUE} and later, manipulating the returned \code{ggplot} class object obtained through the \CRANpkg{ggplot2} package \citep{ggplot22016}.

\subsection{Endpoint selection by means of \texttt{ARE\_tte}}

The ARE gives a criterion to decide whether to use a composite endpoint ($\varepsilon_*$) or to use its more relevant component ($\varepsilon_1$) as the primary endpoint of the trial. Going back to the toy example based on ZODIAC trial, the decision translates into whether to consider death (overall survival, OS) or a composite endpoint (i.e., progression-free survival) as the primary endpoint. 

The ARE method is implemented in the \code{ARE\_tte} function. To compare the two previous mentioned designs, the user can use the following code: 
\begin{example*}
	> ARE\_tte(p0_e1, p0_e2, HR_e1, HR_e2, beta_e1, beta_e2, case, copula, rho, rho_type)
	[1] 9.303        
\end{example*} 

Note that  a value  of $\mathrm{ARE}=9.3>1$  implies that is more efficient considering the composite endpoint (PFS) than the OS as primary endpoint (see the Endpoint Selection subsection). This high efficiency might result in  smaller  required sample size if using PFS instead of OS to attain the same power for a given significance level. In this particular case, a larger proportion of disease progression (DP) events  (0.74 versus 0.59) together with a more marked beneficial effect on DS (0.77 versus 0.91) has lead to an ARE value larger than 1.

\section{CompAREdesign for binary composite endpoints}\label{sect:binary}

\CRANpkg{CompAREdesign} also  includes functions for the design of trials with composite binary endpoints. In this case, composite endpoints are defined as the occurrence of any of the components. Analogously to the time-to-event case, the distribution function for composite binary endpoints relies on the marginal event rates and effect sizes of the composite components, and the correlation between them (see \cite{Bofill2019}).As it is well known, when assessing the efficacy of an intervention against a control treatment based on a binary endpoint  we could use the difference in proportions (or risk difference), the relative risk (or risk ratio), and the  odds ratio. The same applies then for the composite binary endpoints. In \CRANpkg{CompAREdesign}, we implemented the effect size and sample size calculation for trials with composite binary endpoints based on the anticipated values of the composite components and their correlation according to risk difference, relative risk, and odds ratio effect measures. An overview of the functions implemented for binary endpoints can be found in Table \ref{table:compare_functions}. In what follows, we briefly described the two main functions, \texttt{effect\_cbe} and \texttt{samplesize\_cbe}, for the effect size and sample size computation, respectively. 

The function \texttt{effect\_cbe} can be used to compute the effect size of the composite binary endpoint by means of:
\begin{example*}
	> effectsize_cbe(p0_e1, p0_e2, eff_e1, effm_e1, eff_e2, effm_e2, effm_ce ="diff", rho)
\end{example*}
where
\texttt{p0\_e1} and \texttt{p0\_e2}	 denote the probabilities of  $\varepsilon_1$ and $\varepsilon_2$ in the control group, respectively;
\texttt{eff\_e1} and \texttt{eff\_e2} are the anticipated effects for the events $\varepsilon_1$ and $\varepsilon_2$, respectively;
\texttt{rho} is Pearson's correlation between  $\varepsilon_1$ and $\varepsilon_2$.
The effects for  $\varepsilon_1$ and $\varepsilon_2$  can be anticipated in \texttt{eff\_e1} and \texttt{eff\_e2} by means of the difference of proportions, risk ratio, and odds ratio.
The arguments \texttt{effm\_e1} and \texttt{effm\_e2} can be used for specifying the effect measure  preferred.  
Also, using the argument \texttt{effm\_ce}, we specify the effect measure we are interested in for the composite endpoint.

The function \texttt{samplesize\_cbe} can be called by:
\begin{example*}
	> samplesize_cbe(p0_e1, p0_e2, eff_e1, effm_e1, eff_e2, effm_e2, effm_ce ="diff", rho, 
	alpha = 0.05, beta = 0.2, unpooled = TRUE)
\end{example*}
where
\texttt{p0\_e1}, \texttt{p0\_e2}, 
\texttt{eff\_e1}, \texttt{eff\_e2}, and
\texttt{rho} 
\texttt{effm\_e1}, \texttt{effm\_e2} and \texttt{effm\_ce} 
are the parameters explained above; and where 
\texttt{alpha} and \texttt{beta} are the type I and type II errors, respectively;
and \texttt{unpooled} denotes the variance estimate used for the sample size calculation ("TRUE" for unpooled variance estimate, and "FALSE" for pooled variance estimate).

Other functions included in the R package are \texttt{prob\_cbe} to calculate the probability of the composite endpoint, and 
\texttt{ARE\_cbe} to compute the ARE method for binary endpoint, as proposed in  \cite{Bofill2018}.

\section{Simulation feature}\label{sect:sim}

As an additional feature, the package \CRANpkg{CompAREdesign} includes functions to generate data from both the components and the composite endpoint. Using the same input parameters previously seen, the \texttt{simula\_tte} function generates \texttt{samples\_size} times for each treatment arm and the corresponding censoring indicator variables. We assume non-informative administrative right censoring data, thus, all events not  occurring before or  at time $\tau$ (\texttt{followup\_time}) are censored at $\tau$.

Simulations are in general a key tool in the design of RCTs to test the robustness of methods with respect of deviations of the assumptions, sensitivity of sample size calculations regarding the parameter values' assumptions, and compare several methods (e.g., a trialist might wish to compare a non-parametric, a semi-parametric and a parametric method under certain conditions) and/or the efficiency of different trial designs. 

The function returns a data frame with seven columns. The first 6 represent the times (to event or to censoring) and the censoring indicator (\textit{status}) for $\varepsilon_1$, $\varepsilon_2$, and $\varepsilon_*$, respectively. The last column represents the treatment arm. The following example shows simulated data with similar characteristics as those from the ZODIAC trial.

\begin{example}
	> set.seed(12345)
	> rand_data <- simula_tte(p0_e1, p0_e2, HR_e1, HR_e2, beta_e1, beta_e2, case, 
	copula, rho, rho_type, followup_time, sample_size=1000)
	
	> head(rand_data)
	time_e1 status_e1   time_e2 status_e2   time_ce status_ce treated
	1 0.9312005         1 0.1658870         1 0.1658870         1       0
	2 1.0000000         0 0.5646213         1 0.5646213         1       0
	3 1.0000000         0 0.6827438         1 0.6827438         1       0
	4 1.0000000         0 0.7455560         1 0.7455560         1       0
	5 0.1146371         1 0.1954875         1 0.1146371         1       0
	6 1.0000000         0 0.3642334         1 0.3642334         1       0
	
	> tail(rand_data)
	time_e1 status_e1    time_e2 status_e2    time_ce status_ce treated
	1995 0.5836539         1 0.34471237         1 0.34471237         1       1
	1996 1.0000000         0 0.46461461         1 0.46461461         1       1
	1997 1.0000000         0 1.00000000         0 1.00000000         0       1
	1998 0.3927573         1 0.06591404         1 0.06591404         1       1
	1999 0.6078812         1 0.08907998         1 0.08907998         1       1
	2000 0.1594827         1 0.08247508         1 0.08247508         1       1
\end{example}

On the other hand, \texttt{simula\_cbe} simulates two-arm trials with binary composite endpoints.

\section{Discussion} 

Although there are several R packages for the analysis of trials with composite endpoints, to our knowledge, the \CRANpkg{CompAREdesign} package is the first specifically implemented to address the design of  RCTs with composite endpoints. 
The design and analysis of trials with composite endpoints might be specially challenging as they rely on several assumptions of the composite components. For instance, it depends on the functional form of the survival distribution (e.g., the shape parameters of the marginal Weibull distributions) as well as on parameters' values such as the  expected effect sizes, that is to say, the hazard ratios for each composite component. Furthermore, the joint distribution between both component endpoints (including the marginal law and the association between both endpoints) needs to be anticipated. Since anticipation of this long list of arguments might be arduous, \CRANpkg{CompAREdesign} can be a great help in the design of the trial.

The \CRANpkg{CompAREdesign} package (and its corresponding shiny app) may be of particular interest in cases where the precise value of these parameters is unknown. The sensitivity and robustness of results due  to variations in initial values and assumptions can be assessed by means of this package. For instance, by means of the function \texttt{samplesize\_tte}, one could compare several scenarios depending on the shape parameters of the Weibull distribution ($\beta_j$) or the expected effect size in either component ($\mathrm{HR}_j$) as seen in Figure \ref{Figure_samplesize}. In addition, \CRANpkg{CompAREdesign} includes several association measures and copula functions for addressing different dependency structures between the time to each of the composite components. Although the choice of the copula is not straightforward, preliminary simulation results \citep{Cortes2019} show a small impact of the choice of the copula on the corresponding percentiles for the composite endpoint, and consequently on the decisions one might have to take.

\CRANpkg{CompAREdesign} has implemented  Spearman's correlation and Kendall's tau as association measures. The former is much more frequent and is the one we recommend for time-to-event studies. 

To compute the sample size the user can choose between Schoenfeld's and Freedman's formulas. The choice should take into account that their appropriateness   depends on the deviation from the proportional hazards assumption: the greater the deviation (less constant hazard ratio over time), the better to use  Freedman's instead of Schoenfeld's formula \citep{Friedman1996}.

This package is restricted to two-arm RCTs. As future work, we consider expanding the R package to trials with more than two arms. In particular, we plan to extend the package to also include multi-arm trials where the efficacy of $K$ treatments is tested against a shared control \citep{Jaki2019}. Also, the package could be expanded to include some adaptive features, such as sample size reassessment \citep{Bofill2022} and treatment selection \citep{Bretz2009}. Finally, the package could be enhanced to include trials with composite endpoints with more than two components. 

%%%%%%%%%%%%%%%%%%%%%%%%%%%%%%%%%%%%%%%%%%%%%%
%%                                          %%
%% Backmatter begins here                   %%
%%                                          %%
%%%%%%%%%%%%%%%%%%%%%%%%%%%%%%%%%%%%%%%%%%%%%%

\subsubsection*{Funding}%% if any

This work was supported by the Ministerio de Economía y Competitividad (Spain) under Grant PID2019-104830RB-I00 and the Departament d'Empresa i Coneixement de la Generalitat de Catalunya (Spain) under Grant 2017 SGR 622 (GRBIO). Marta Bofill Roig is a member of the EU Patient Centric Clinical Trial Platforms (EU-PEARL). This Joint Undertaking receives support from the European Union’s Horizon 2020 research and innovation program and EFPIA and Children’s Tumor Foundation, Global Alliance for TB Drug Development nonprofit organization, Springworks Therapeutics Inc. This publication reflects the authors' views. Neither IMI nor the European Union, EFPIA, nor any Associated Partners are responsible for any use that may be made of the information contained herein.
%%%%%%%%%%%%%%%%%%%%%%%%%%%%%%%%%%%%%%%%%%%%%%%%%%%%%%%%%%%%%%%%%%%%%%%%%%
%%%%%%%%%%%%%%%%%%%%%%%%%%%%%%%%%%%%%%%%%%%%%%%%%%%%%%%%%%%%%%%%%%%%%%%%%%
%%%%%%%%%%%%%%%%%%%%%%%%%%%%%%%%%%%%%%%%%%%%%%%%%%%%%%%%%%%%%%%%%%%%%%%%%%

%\newpage
%
%\footnotesize
%%\bibliography{references}
%%\bibliographystyle{dinat}
\bibliography{CortesBofillGomez} 
%%\printbibliography
%\clearpage

%%%%%%%%%%%%%%%%%%%%%%%%%%%%%%%%%%%%%%%%%%%%%%%%%%%%%%%%%%%%%%%%%%%%%%%%%%
%%%%%%%%%%%%%%%%%%%%%%%%%%%%%%%%%%%%%%%%%%%%%%%%%%%%%%%%%%%%%%%%%%%%%%%%%%
%%%%%%%%%%%%%%%%%%%%%%%%%%%%%%%%%%%%%%%%%%%%%%%%%%%%%%%%%%%%%%%%%%%%%%%%%%

\clearpage

\newpage

%\includepdf[pages=-]{Supp\_material.pdf}
%\includepdf[pages=-]{NCCsuppmat.pdf}

\clearpage

\end{document}